# Hybrid electrostatic-piezo MEMS photonic integrated modulators

Thuy-Linh Le,[1,*] Hardit Singh,[1,*] Julia M. Boyle,[1] Matthew Zimmermann,[1] Andrew J. Leenheer,[2] Daniel Dominguez,[2] Matt Eichenfield,[2,3,4] and Mark Dong[1,5]

[1]*The MITRE Corporation, 202 Burlington Road, Bedford, Massachusetts 01730, USA*
[2]*Sandia National Laboratories, P.O. Box 5800 Albuquerque, New Mexico, 87185, USA*
[3]*College of Optical Sciences, University of Arizona, Tucson, Arizona 85719, USA*
[4]*Electrical, Computer, and Energy Engineering, University of Colorado Boulder, Boulder, Colorado 80309, USA*
[5]*mdong@mitre.org*
[*]*Equal contributors*

**Abstract:** Programmable photonic integrated circuits (PICs) have recently emerged as an important technology for quantum information science and artificial neural networks. In particular, PICs with MEMS-based modulators have the advantages of voltage-based control, ultra-low-energy consumption, cryogenic compatibility, and CMOS-foundry support. Here we report a cantilever optical modulator that utilizes hybrid piezoelectric and electrostatic tuning forces together on a monolithic silicon nitride (SiN) PIC platform. The device achieves actuation of visible-wavelength light with quasi-static tuning up to 10 kHz at 1.5 $V_\pi$-cm as well as high-speed (>20 MHz) AC modulation with dynamically adjustable (25 - 40 MHz) mechanical resonances. We report the physics of how geometric nonlinearities such as capacitive pull-in give rise to suspended and contacted cantilever modes. These reversible operating regimes generate different strain profiles and boundary conditions which are responsible for the active tuning of the mechanical resonances. Our proof-of-concept electrostatic-piezo modulator shows promising potential in large-scale programmable PICs applied to high-speed optical switching and optomechanical sensing.

# 1. Introduction

Photonic integrated circuits (PICs) [1] are emerging as a versatile and promising technology enabling the manipulation of light for a wide variety of applications, all in a chip-scale package. Some recent demonstrations include addressing qubits for quantum information processing [2-4] or performing efficient optical computation as part of artificial neural networks [5-6]. Current large-scale programmable PICs are accessible at the foundry-level designed under various architectures [7-8], typically using thermo-optic modulation [9-12] or microelectromechanical (MEMS)-based modulation [13-17] on Si substrates. Programmable optical phase shifts with MEMS modulators induce strain or displacement [13-18] of materials to change modal refractive indices or on-chip interferometer path lengths to perform various switching or unitary operations, resulting in high switching speeds [19-20] and very low static power consumption [21]. Therefore, the advantages of MEMS actuation are highly suitable for AI and quantum applications where energy efficiency or cryogenic compatibility are critical requirements.

MEMS optical modulation is commonly implemented with two approaches. Electrostatic actuation is typical in MEMS-photonics foundries using processes in silicon [13], making reliable modulators (<1 MHz bandwidth) at scale predominantly at telecom wavelengths with a few successful demonstrations in the visible regime [22-23]. Alternatively, piezoelectric actuation [24] operates faster (>100 MHz) and often requires additional custom post-processing to integrate piezoelectric materials [25-27], although some foundry support has recently emerged [20] [28]. There is an open question of whether simultaneously incorporating both actuation forces at the foundry level would improve MEMS modulators at visible wavelengths, with the physical interplay between these two mechanisms not yet explored for integrated photonics. Further developing reliable, foundry-processed optical MEMS technology and investigating their operation in the visible-wavelength regime would be impactful for important applications of programmable PICs.

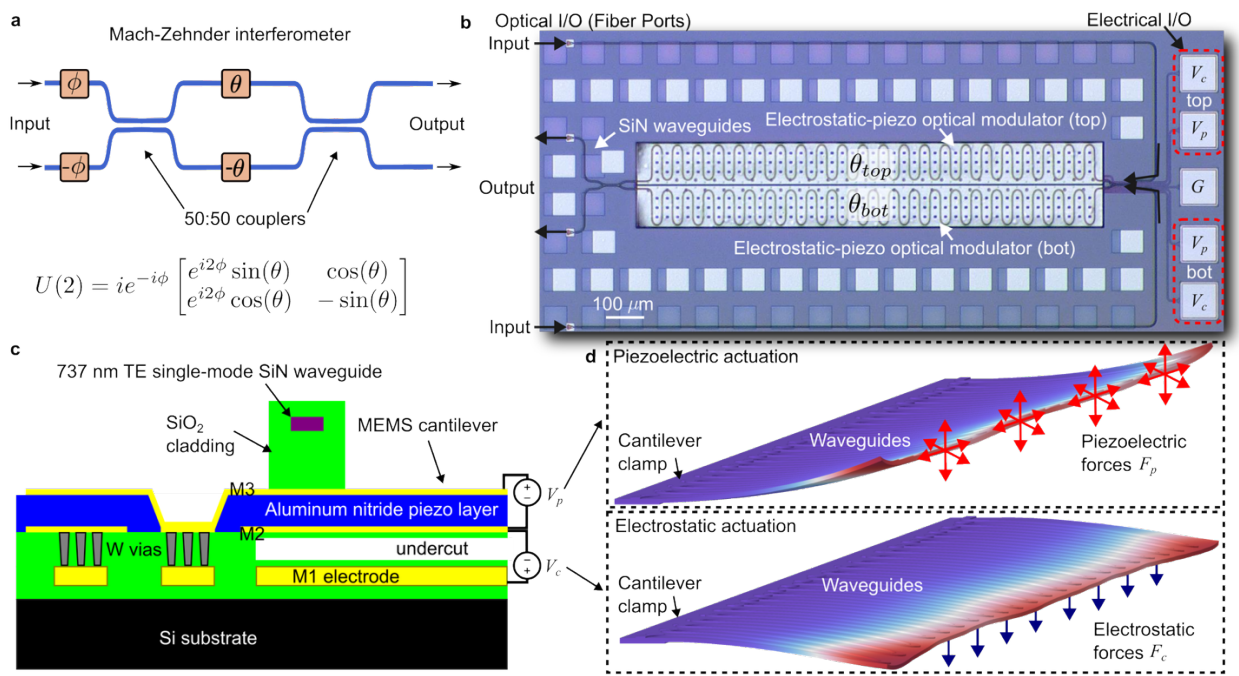

**Figure 1: Hybrid electrostatic-piezo MEMS photonic integrated modulators.** a) A basic programmable Mach-Zehnder diagram with tunable phases $\phi, \theta$ implementing the unitary operation $U(2)$ b) Optical microscope image of the hybrid electrostatic-piezo optical modulator with labeled optical I/O (fiber ports) and electrical I/O (probed pads) c) Cross section cartoon of the full layer stack showing applied piezo and capacitive voltages $V_p, V_c$ respectively d) Diagram showing deformed geometries of the cantilever and optical waveguides under application of piezoelectric and electrostatic forces.

Here we report a novel type of optical MEMS modulator suitable for reconfigurable PICs that integrates both piezoelectric and electrostatic actuation on a monolithic visible-wavelength PIC layer stack. The modulator builds off our prior piezoelectric PIC platform [19] fabricated with CMOS-compatible processes. By modifying the interconnects between existing metal layers, we integrated a parallel-plate capacitor beneath the existing piezoelectric layers without substantially changing the fabrication process, thus enabling reconfigurable phase shifters with an additional electrostatic degree of freedom. The concept illustrated in Fig. 1a, b shows the basic programmable Mach-Zehnder interferometer and an optical image of the fabricated device consisting of two electrostatic-piezo cantilevers, one per arm. The optical and electrical input / output (I/O) ports enable independent voltage-based actuation of either cantilever. The cross section of the photonic layer stack, shown in Fig. 1c, consists of the M1 metal layer used for signal routing as well as the positive voltage electrode of the capacitive tuner. The stack continues with insulating oxide layers and an undercut layer made of sacrificial amorphous Si (a-Si) whose thicknesses define the capacitor, followed by the M2 and M3 metal layers sandwiching the aluminum nitride (AlN) piezo layer, and several tungsten vias (W). Finally, a bottom oxide cladding separates the SiN waveguiding layer, which is capped off with a top oxide cladding. The combination of the M1, M2, and M3 electrodes enables application of compressive and tensile piezoelectric forces along with unipolar parallel-plate electrostatic forces (Fig. 1d) to the cantilever structure. Average layer thicknesses, including insulating oxide layers, are listed in Table 1.

Table 1: Approximate layer thicknesses of the electrostatic-piezo integrated photonics platform

| Layer | M1 (Al) | SiO$_2$ below a-Si | a-Si undercut | SiO$_2$ above a-Si | M2 (Al) | AlN piezo | M3 (Al) | SiO$_2$ bottom cladding | SiN | SiO$_2$ top cladding |
|---|---|---|---|---|---|---|---|---|---|---|
| Thickness | 850 nm | 400 nm | 200 nm | 200 nm | 120 nm | 450 nm | 250 nm | 750 nm | 300 nm | 300 nm |

## 2. Electrostatic-piezo modulator design and characterization

The basic concept of the hybrid cantilever modulator follows our previous design [18] in which waveguides are meandered along the overhang of the cantilever. The main device we characterized for this paper consists of two independently controlled cantilevers, each with a 91-μm overhang and a 1.292 mm width (with additional device designs characterized in Supplement S2). The total waveguide propagation length is 4.71 mm per cantilever plus two ~50:50 evanescent directional couplers to complete the MZI, with total per MZI losses estimated around ~2 dB. Finally, angled on-chip grating couplers for mode-matching to a fiber mode complete the device design.

We experimentally characterized the modulator performance using a 737 nm wavelength laser delivered via a 4-port fiber array and detected with standard photodiodes. We generated various voltage signals applied to the top or bottom cantilever electrical pads, labeled $V_p$ and $V_c$ respectively (Fig. 1b), enabling control of the piezoelectric and electrostatic forces, respectively. For DC characterization, we applied a 100 Hz ramp through a 20x gain, high voltage amplifier. We characterized the piezoelectric layer with a ramp from -40 to +40 V and the capacitive layer with a ramping voltage from 0 to 80 V, independently measuring either the top or the bottom cantilever. For AC characterization, a 50% duty-cycle square wave with a DC bias was generated with different characteristics, depending on if it was applied to the piezo or capacitive actuator. The piezoelectric signal was a square with a 10 V peak-to-peak signal and a 100 kHz repetition rate that skips the 20x amplifier due to bandwidth constraints. The capacitive signal was a square with a 3.18 V peak-to-peak signal, 38.2 V DC offset, and a 1 kHz repetition rate.

Fig. 2 shows the results of the basic actuation experiment. We observed the piezoelectric switching behavior of a single-arm of the MZI (Fig. 2a, b were tested on the bottom cantilever), demonstrating linear actuation with a fitted $V_\pi$ of ~26V, which corresponds to 12 $V_\pi$-cm, and high-speed switching times of ~32 ns. These results match well to previously reported piezoelectric modulators [18]. In contrast, the electrostatic actuation (Fig. 2c) shows nonlinear modulation where the cantilever passes through three distinct regimes: suspended, pinned, and clamped corresponding to clamped-free, clamped-pinned, and clamped-clamped boundary conditions for suspended beam [29]. We note the most efficient switching behavior occurs in the pinned and pinned-to-clamped transition regime, where a π phase shift can be obtained with only ~3.2 V swing. A square-wave AC switching experiment, DC biased in the pinned regime, is plotted in Fig. 2d where the electrostatic rise and fall times (~59 μs and ~40 μs, respectively) are about three orders of magnitude slower than those of piezoelectric. The shape and bandwidth of the electrostatic curves are also similar to previously reported work [21]. The difference in modulation response speed shows that, despite the same cantilever structure and compliance, piezoelectric and strain propagation response times are much faster than deformations induced by electrostatic forces. We note the Fig. 2 experiments were performed with the piezo and electrostatic control signals applied separately – more information on the experimental methods is given in Supplement S1.

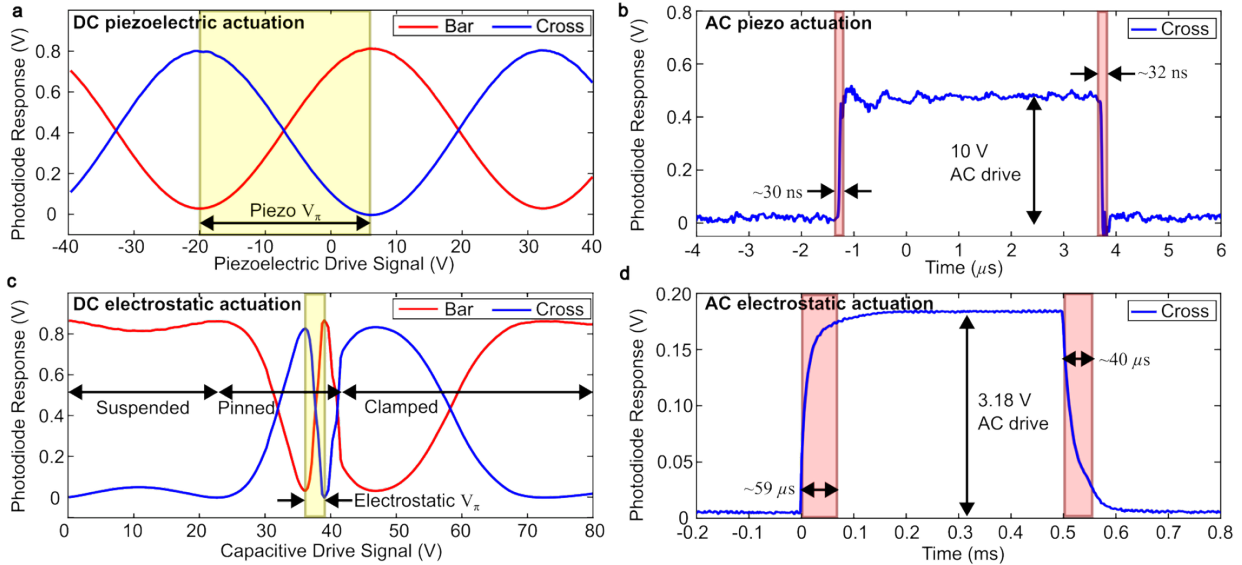

**Figure 2: DC and AC characterization of piezoelectric and electrostatic actuation.** a) Photodiode response of bar and cross MZI outputs under single-arm quasi-DC (100 Hz) piezoelectric actuation with labeled $V_\pi$ of ~26 V b) Photodiode response of single-arm piezoelectric actuation under an applied 10 V peak-to-peak AC square wave, showing rise and fall times of ~30 – 32 ns c) Photodiode response of bar and cross MZI outputs under single-arm quasi-DC (100 Hz) electrostatic actuation with a labeled switching $V_\pi$ of ~3.2 V and three cantilever operating regimes d) Photodiode response of single-arm electrostatic actuation under an applied 3.18 V peak-to-peak AC square wave mixed with a DC bias of 38.2 V. Electrostatic switching shows longer rise and fall times of ~59 μs and ~40 μs respectively, compared to those of piezoelectric switching.

We proceed to investigate the interplay between the two acting forces. We first measured the modulator's small-signal frequency response when probed piezoelectrically with varying DC capacitive voltages $V_c$, shown in Fig. 3a. At each $V_c$, a 6 V peak-to-peak AC drive with swept frequency was applied after biasing the MZI to the 50:50 point with the unused (top) cantilever. We observe several lower order peaks from 100 kHz to 10 MHz frequency and a clear main peak around ~23 MHz, which we define as the "primary mode," i.e. the mode that contributes most to the optical phase shift. This primary mode exhibits dynamical shifting behavior to higher frequencies as more capacitive voltage is applied, especially increasing sharply after $V_c$ ~25 V (Fig. 3b). The dynamical frequency shift is likewise accompanied by a corresponding decrease in actuation enhancement (Fig. 3c shows an initial quality factor by Lorentzian fit of ~15) such that by $V_c$ = 80 V any enhancement is no longer visible. We also observe a secondary, higher order peak appearing around ~65 MHz that follows a similar dynamical tuning pattern with $V_c$.

To understand the primary mode's higher resonance enhancement over other lower-order modes, we performed finite-element simulations (COMSOL Multiphysics) of the mechanical eigenmodes for our cantilever layer stack and geometry, shown in Fig. 3d. While we found many eigenmodes, there was a class of modes around ~22 MHz whose displacements along the length of the cantilever roughly matched the periodicity of the waveguide meanders, as seen in the top and front views of Fig. 3d (plotted displacements are magnified for visual clarity, not to scale). We identified this type of mode as the primary mode measured in the experiment due to its additive nature of the phase shifts across the entire length of the cantilever.

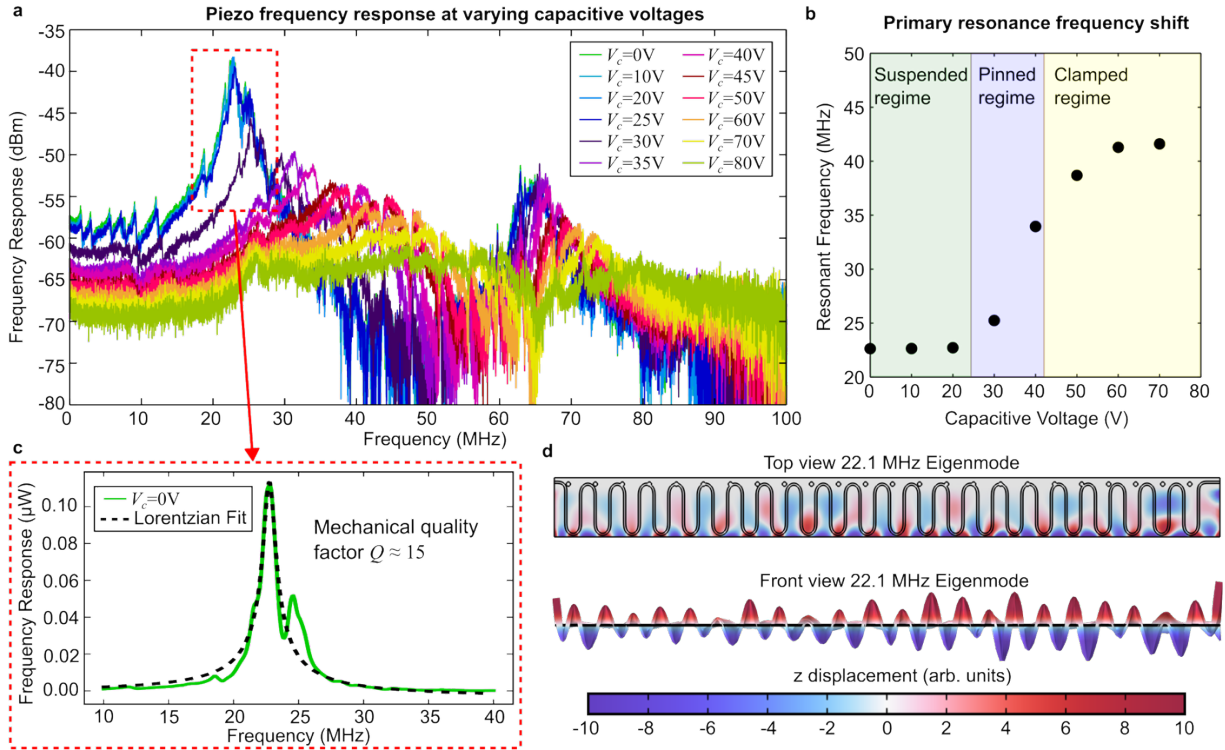

**Figure 3: Piezoelectric frequency response vs capacitor voltage.** a) Measured frequency response of the cantilever when driven with piezoelectric forces at varying applied capacitive voltages. One large peak (the primary mode) around 23 MHz emerges as the mode most effectively contributing to the optical phase shift. This primary mode shifts to higher frequency and decreases in strength as capacitive voltage increases b) Extracted resonance frequency of the primary mode as a function of applied capacitive voltage with regime overlays c) Fitted Lorentzian to the primary mode showing a mechanical quality factor $Q$ of ~15 when the capacitor is unbiased d) 3D eigenmode simulation of the cantilever geometry using finite-element software showing a ~22.1 MHz mechanical mode (displacements are greatly magnified for visual clarity). This mode resembles the primary mode that is of interest due to the periodicity across the length of the cantilever matching those of the waveguide meanders.

We note the polarity of the cantilever's frequency tuning (Fig. 3b) initially appears opposite of what is expected in similar geometries. For example, non-contact atomic force microscopy (AFM) shows mechanical resonance frequencies decrease as van der Waals forces get stronger from a smaller separation gap between the cantilever and the scanned surface [30-31]. However, the primary mode (Fig. 3d) is a higher order 3D flexural mode [32] whose deflection is not accurately captured by 1D beam theory. Moreover, the small initial airgaps (200 nm) in our released structure would also easily allow the cantilever to transition to geometrically nonlinear regimes as observed in Fig. 2c. We explore these features in the following section.

## 3. Device theory and operating regimes

Following previous work [18], the physical phase shift mechanism of these devices is dominated by axial strain imparted to the waveguides, thus resulting in optical path length changes. While piezoelectric materials embedded in the cantilever will directly produce axial strain, the electrostatic forces instead act externally on the suspended beam, pulling it into different geometric regimes. We explore these regimes (labeled suspended, pinned, and clamped) and how piezoelectric and electrostatic actuation are affected.

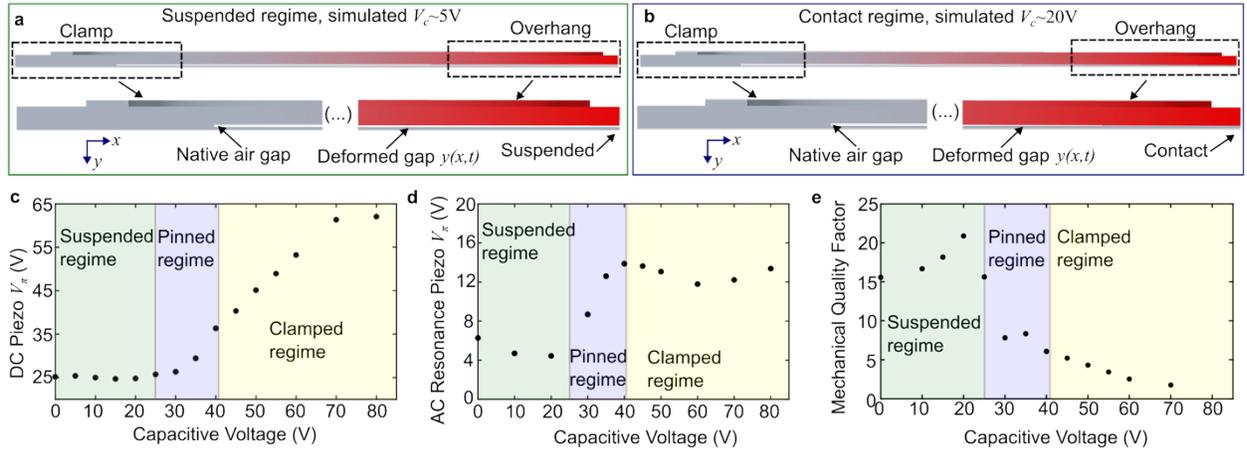

**Figure 4: Cantilever suspended and contact regimes.** a) 3D finite-element mechanical displacement simulation of cantilever geometry with a 5 V applied capacitor voltage showing the native air gap (by the clamp) and a decreased air gap (at the overhang) while remaining suspended b) 3D finite-element mechanical displacement simulation of cantilever geometry with a 20 V applied capacitor voltage showing the overhang has made contact into the bottom surface c) Measured and fitted quasi-DC piezo $V_\pi$ at varying capacitive voltages d) Measured and fitted piezo $V_\pi$ when driven at the primary mode's resonance frequency at varying capacitive voltages e) Extracted mechanical quality factor of the primary mechanical mode at varying capacitive voltages.

We performed additional structural mechanics simulations using COMSOL with the layer stack and geometry from Fig. 1c to investigate the mechanical behavior of the cantilever modulator. Assuming a flat cantilever as the initial condition, the static deformation was calculated at increasing potential differences ($V_c$) set between the cantilever and the bottom plate. The simulation shows the cantilever remains suspended throughout its length at $V_c = 5$ V but contacts the bottom plate at $V_c \sim 20$ V (Fig. 4a, b respectively) when the cantilever edge has moved the vertical separation of the air gap (200 nm). Therefore, we expect the device to transition to contact boundary conditions and exhibit distinct modulation characteristics for both electrostatic and piezo actuation.

In addition to resonance frequency shift, we measured the DC piezo $V_\pi$, AC piezo $V_\pi$, and extracted the mechanical quality factor at varying applied capacitive voltages. We observe an inflection point (Fig. 4c and 4d) where both the DC and AC $V_\pi$ starts increasing past around $V_c \sim 25$ V, which reasonably matches the simulations (Fig. 4a, b) given the fabricated cantilever is not perfectly flat when released due to residual axial strain in the layer stack. We identify the $V_c <25$ V regime as the suspended regime and the $V_c >25$ V as the "contact regime" for this device, which consists of a pinned (25 V - ~40 V) and then a clamped geometry (>40 V) matching the labels in Fig. 2c. The DC $V_\pi$ increases linearly well into the clamped section and tapers around 70 V, while the AC $V_\pi$ increases rapidly at the pinned transition and tapers off sooner at the pinned-clamped transition. The behaviors for both AC and DC indicate that $V_\pi$ increases from additional mechanical losses from contacting the bottom substrate. Lastly, we performed Lorentzian parametric fits for the frequency response data from Fig. 3a at each capacitive voltage. The mechanical $Q$ is plotted in Fig. 4e. Notably, for capacitive voltages from 0 to 20 V the $Q$ shows a small increase (with a corresponding decrease in the AC $V_\pi$ in Fig. 4d) while for capacitive voltages beyond ~40 V the $Q$ monotonically decreases. We note that the initial increasing $Q$ can be explained by energy being elastically stored as tensile strain in the cantilever, which contributes to reduced energy dissipation [33]. Additional detailed data on the $V_\pi$ measurements and Lorentzian fits are given in Supplement S1.

Building off the understanding of the distinct operating regimes, we describe the physics of the dynamical tuning of the resonance frequency as resulting from four distinct effects: 1) a change in the strength of the external electrostatic forces 2) a change in the axial strain whether tensile or compressive of the cantilever beam [29] [34]; 3) boundary condition changes [35] due to contact; and 4) effective cantilever length changes post-clamp [36]. Our experimental results suggest that, while we may have some frequency shift due the first two effects, (similar to the case with van der Waals forces) the largest resonance frequency shifts only occur after contact. This explains the increasing resonance frequency (as opposed to decreasing) from both effects 3 and 4: boundary condition changes from fixed-free to fixed-pinned or fixed-fixed will make the beam effectively much stiffer and additional capacitive pull-in effects only shortens the overall cantilever length.

We further confirmed the suspended and contact regimes with additional white light profilometer measurements of the cantilever surface as well as verifying the buckling effect of the structure post-clamp. We applied varying $V_c$ and measured topographical deformations in Fig. 5. We observed that the cantilever's rest state at $V_c = 0$ V (Fig. 5a) is slightly curled up and appears to be suspended. However, as $V_c$ increases, the cantilever's edge collides with the bottom substrate and eventually flattens significantly. We also plot a 1D cross-section (Fig. 5b) for the measured vertical displacement across the length of the cantilever. The cantilever's edge moves downwards vertically until stopping around ~30 V, indicating a likely collision. After a threshold voltage between 30V and 40V, the cantilever's main body snaps down and flattens, confirming the pinned-to-clamped transition. At capacitive voltages beyond 40 V, the cantilever's effective length decreases as more of the material becomes flattened on the bottom substrate. Despite geometrically nonlinear buckling and contact effects, setting $V_c = 0$ V returns the cantilever to its rest state with minimal hysteresis (Fig. 5c). This allows for reliable control of the device for repeatable transitions between the separate regimes. We note the oxide ridge height appears inverted due to how the profilometer processes reflected light from transparent but higher refractive index materials as a larger distance traveled, so the vertical offsets at these locations will not be accurate. However, this limitation does not significantly affect the overhang region, which is our main focus.

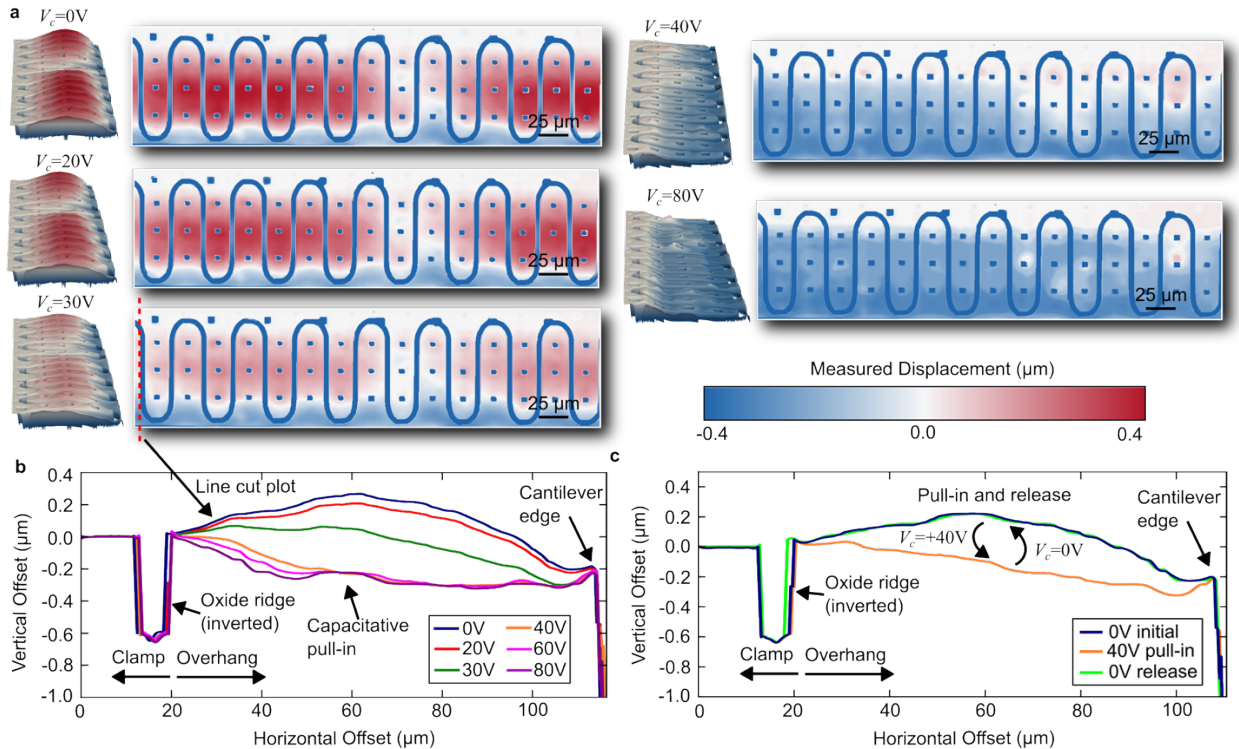

**Figure 5: White-light profilometer measurements of cantilever deformation.** a) 3D plot of partial side view and top view of measured cantilever height at varying applied capacitive voltages. A transition occurs between 20 V - 30 V showing contact and then capacitive pull-in effects beyond 30 V. b) 1D line cut plot of measured profilometer data illustrating the pull-in transition. We note the oxide ridge's vertical offset data is inverted. c) 1D line cut plot of measured profilometer data cycling between 0V and 40V applied capacitive voltages showing quick release with minor hysteresis.

The profilometer measurements confirm our previous insights about the electrostatic actuation mechanisms. Specifically, quasi-static electrostatic tuning (Fig. 2c) exhibits minimal phase shifting until ~25 V in which the electrostatic tuning starts to become significant (suspended regime). Then a highly sensitive tuning region occurs indicating large axial strains generated from contact, up to ~40 V (pinned regime) after which the tuning sensitivity starts to decrease (clamped regime). Similarly, the primary mechanical mode's resonance shift (Fig. 3b) is most sensitive during the 30 – 40 V regime, suggesting the boundary condition change is the most responsible followed by further effective overhang decreases (between 40 V and 80 V). The flattening effect also explains the additional mechanical loss pathways for piezoelectric actuation (Fig. 4c-e) as capacitive voltage is increased.

Lastly, we comment that practical operation of these devices will depend on the application. The electrostatic actuation and tuning of the mechanical resonance are most sensitive right around the pinned regime. For resonantly actuated switches [37] or setting a state for unitary operations [38], this operating point offers the lowest electrostatic $V_\pi$ and a few MHz mechanical tuning without significantly degrading piezo operation (Fig. 4). However, for optomechanical sensing [39-40], biasing the cantilever to a specific mechanical frequency of interest or to the edge of pull-in may allow for a more sensitive device.

## 4. Discussion

We presented and characterized a hybrid electrostatic-piezo integrated optical modulator for visible wavelength actuation. This device allows for flexible DC or AC operation, enabling 1.5 $V_\pi$-cm DC actuation at 10 kHz capacitive switching but >20 MHz AC piezoelectric actuation for both low hold-power capacitive tuning and high-speed switching. These results advance the technology for foundry processed MEMS photonics and visible-wavelength, cryogenically compatible PICs with important applications in quantum information, optical computing for AI, and optomechanical sensing.

Several possible device improvements may be considered for future work. We characterized several cantilevers of the same design in addition to two other designs for fabrication yield, performance differences, and overall reliability (see Supplement S2). The yield for the reported device design (Device C in the Supplement) was ~84%, with the primary failure mode being not the mechanical or optical components but some electrical vias unable to handle repeated application of >80 V. Reducing the oxide thicknesses in between the capacitor plates would improve electrostatic strength thus lowering required $V_c$, although careful attention needs to be paid to the initial cantilever curling by balancing the stresses underneath and above the piezo layer upon release [41] due to its strong dependence on the overhang length and the thickness of the $SiO_2$ above the a-Si layer. By carefully tailoring the cantilever initial flatness by changing the length of the overhang and $SiO_2$ layer thicknesses, the suspended-to-pinned transition may be accessed with much lower bias voltage. Further investigation of the cantilever cross-sectional area may also improve axial strain generation (such as transforming a cantilever to a slender beam to facilitate buckling) or etching additional holes to reduce mechanical stiffness depending on the desired resonance frequency range. These efforts should improve overall device footprint and actuation efficiency.

**Funding.** MITRE Quantum Moonshot Project; NSF FuSe 2 Award ECCS-2425611; U.S. Department of Energy, Office of Science

**Acknowledgments.** M.D. thanks M. Saha, A. Witte, and K. Palm for helpful technical discussions. M.D. also thanks D. Englund, G. Gilbert, and R. Han for their support. Sandia National Laboratories is a multimission laboratory managed and operated by National Technology & Engineering Solutions of Sandia, LLC, a wholly owned subsidiary of Honeywell International Inc., for the U.S. Department of Energy's National Nuclear Security Administration under contract DE-NA0003525. This paper describes objective technical results and analysis. Any subjective views or opinions that might be expressed in the paper do not necessarily represent the views of the U.S. Department of Energy or the United States Government.

**Disclosures.** The authors declare no conflicts of interest.

**Data availability.** Data underlying the results presented in this paper are not publicly available at this time but may be obtained from the corresponding author upon reasonable request.

**Supplemental document.** See Supplement for supporting content.

# Hybrid electrostatic-piezo MEMS photonic integrated modulators: Supplement

Thuy-Linh Le,[1,*] Hardit Singh,[1,*] Julia M. Boyle,[1] Matthew Zimmermann,[1] Andrew J. Leenheer,[2] Daniel Dominguez,[2] Matt Eichenfield,[2,3,4] and Mark Dong[1,5]

[1]*The MITRE Corporation, 202 Burlington Road, Bedford, Massachusetts 01730, USA*
[2]*Sandia National Laboratories, P.O. Box 5800 Albuquerque, New Mexico, 87185, USA*
[3]*College of Optical Sciences, University of Arizona, Tucson, Arizona 85719, USA*
[4]*Electrical, Computer, and Energy Engineering, University of Colorado Boulder, Boulder, Colorado 80309, USA*
[5]*mdong@mitre.org*
[*]*Equal contributors*

## S1. Additional experimental details

Here we present some additional information on our experimental data and characterization methods. For characterization plots in Fig. 2 of the main text, we only illuminated one input of the MZI at a time while collecting from both outputs. We optimized the input laser polarization via a manual three paddle polarization controller to ensure the incoming light was TE polarized by maximizing the throughput power to match the grating's TE coupling preference. Electrical control voltages were delivered via an RF probe touching down on the electrical pads. On the detection side, we used photodiodes whose signals were connected to input channels of a FPGA-based digitizer (Liquid Instruments MOKU:Pro) for recording the signal. For active actuation, electrical signals were generated from an arbitrary waveform generator (AWG) (also the MOKU:Pro) and a 20x gain amplifier. The unused cantilever was sometimes used to statically bias the MZI phase depending on the experiment or simply grounded. Based on electrical impedance measurements, we estimate the electrostatic and piezoelectric capacitances to be ~9 pF and ~26 pF, respectively.

For Fig. 4 of the main text, the DC $V_\pi$ measurements were performed with a 100 Hz, 4 V peak-to-peak ramp. Fig. S1 shows sample traces of how the measurement data were processed.

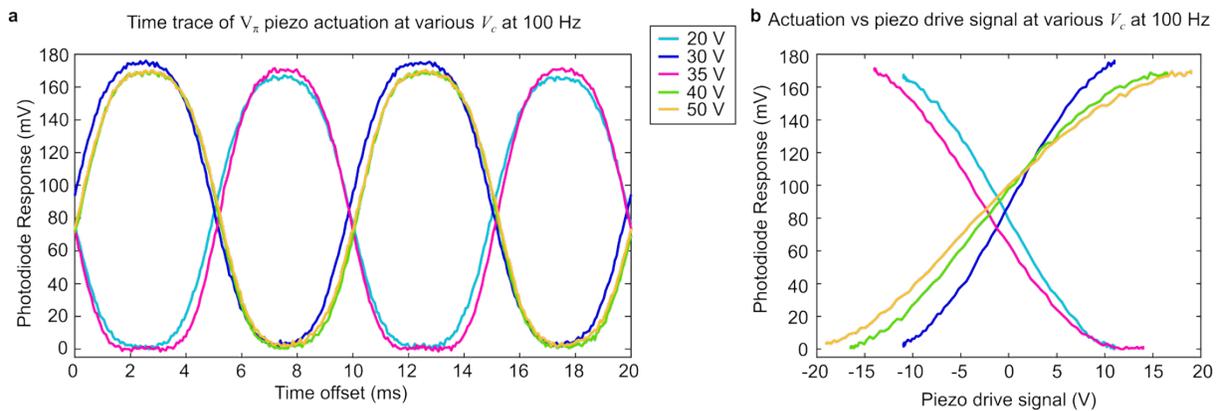

**Figure S1: Time-domain traces and fits of DC piezoelectric $V_\pi$.** a) Plots of quasi-DC (100 Hz) sweep in time domain at varying capacitive voltages. b) The same actuation replotted against the piezo drive signal. A fitted sinusoid to the DC actuation was used to determine the DC $V_\pi$ seen in Fig. 4 in the main text.

The AC $V_\pi$ measurements were performed similarly with a varying frequency matching that of the primary mode, tracking the resulting peak-to-peak sinusoid and fitting the output. For these traces, there is often a phase offset between the modulated output and the input sine due to resonance-induced phase shifts. Here we separately fitted each trace to extract the AC $V_\pi$. Fig. S2 shows results for a few capacitive voltages.

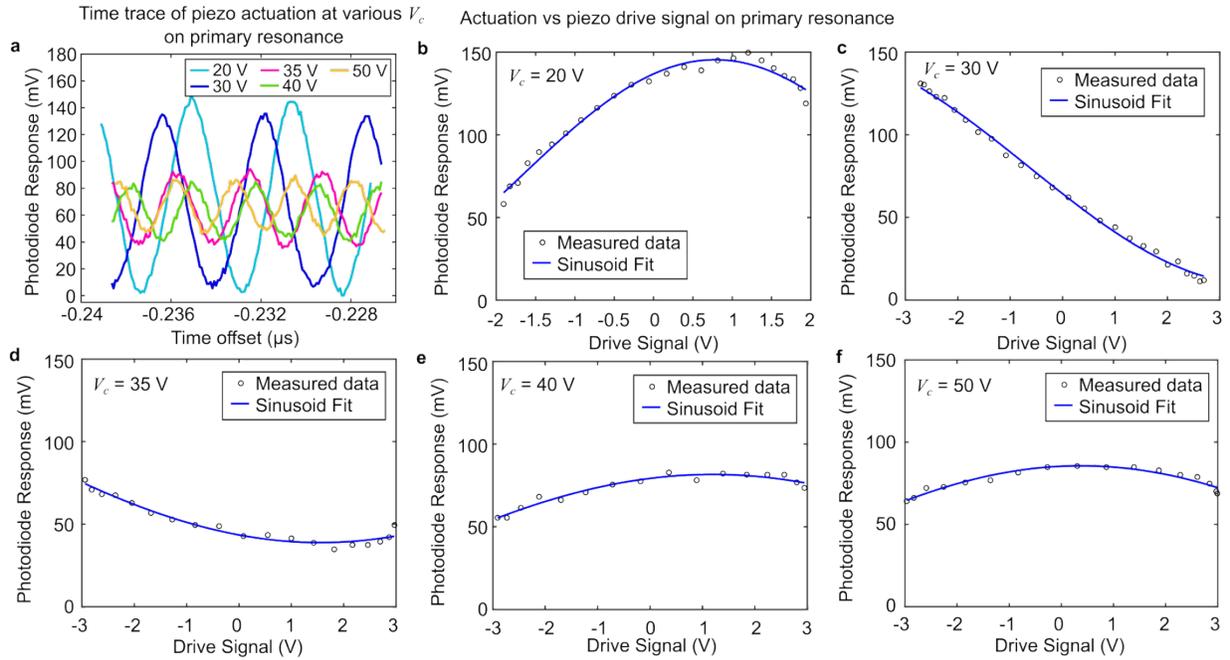

**Figure S2: Time-domain traces and fits of AC piezoelectric $V_\pi$.** Plots of AC actuation whose frequency matches that of the primary mode for an applied capacitive voltage $V_c$. a) Time trace of resonant actuation of the primary mode at varying capacitive voltages, showing a clear decrease in actuation strength as $V_c$ is increased. b)-f) The replotted actuation vs drive signal voltage showing raw data and fitted sinusoid for extraction of the AC $V_\pi$ seen in Fig. 4 in the main text.

For data in Fig. 4e of Lorentzian fits of the frequency response peaks we applied locally weighted scatter plot smoothing (python LOWESS) to help facilitate the fitting, albeit the resulting fits were not substantially different from simply fitting the raw data. The mechanical $Q$ was calculated using the ratio of the frequency peak with the full width at half maximum. Fig. S3 below shows some of the fits performed. The fitting generally became more difficult as the resonator quality factor decreased from excessive losses after the contact (the $V_c$ = 80 V Lorentzian fit did not converge and was not included).

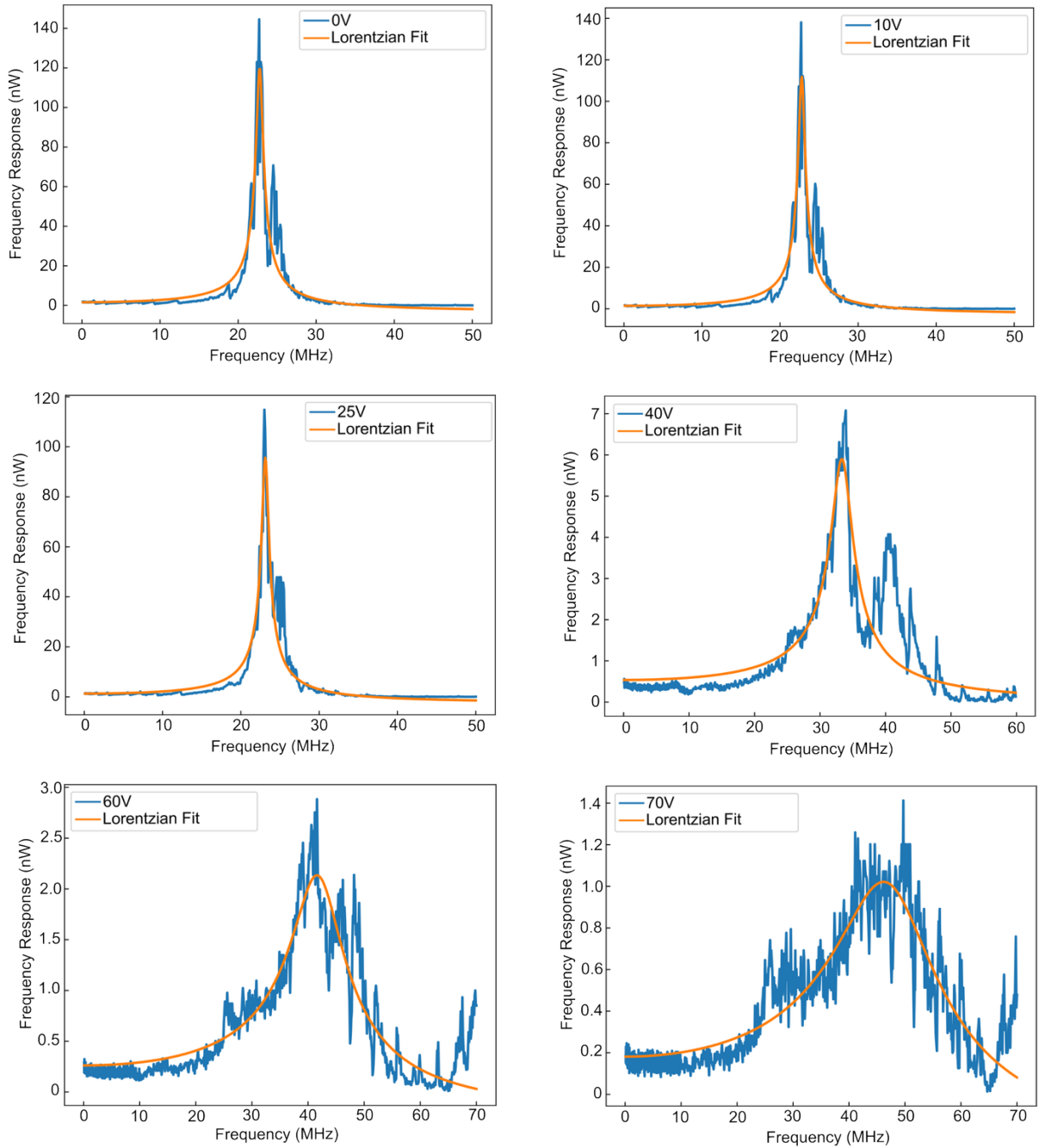

**Figure S3: Lorentzian fits for frequency response data from Fig. 3a at representative capacitive voltages.** As the capacitive voltage is increased, the peak response decreased and the width of the resonance became larger. This corresponds to the decreasing Q values presented in Fig. 4e.

## S2. Additional device characterization data

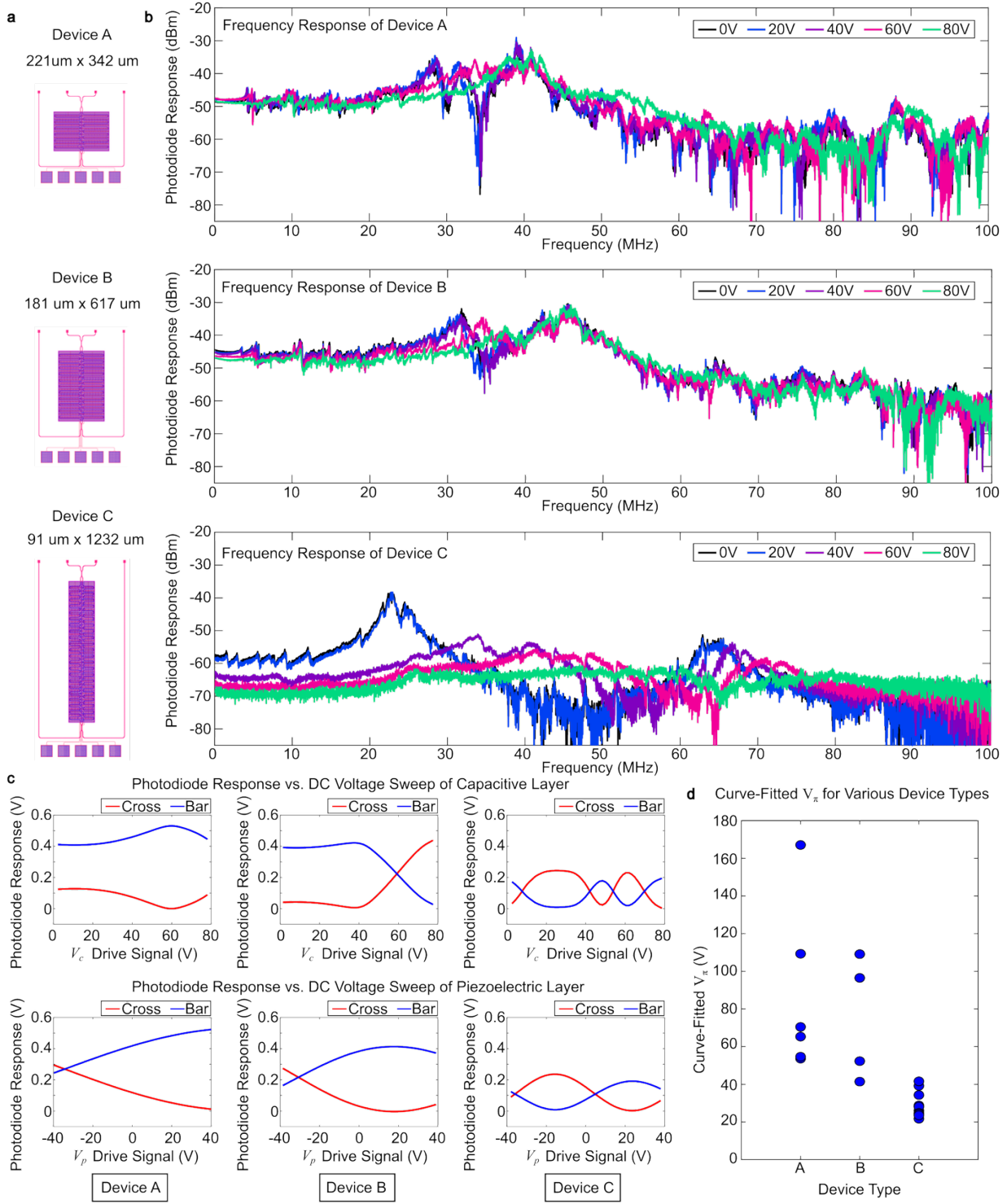

**Figure S4: Comparison of Device Type A, B, and C.** a) GDS II top-view of the layout of the 3 device types b) frequency response comparison between the 3 devices in response to capacitive tuning c) side-by-side comparison of the photodiode response quantifying the electrostatic response to piezoelectric and capacitive layer drive signals d) the curve-fitted piezo $V_\pi$ for 23 devices (6 - Type A, 6 - Type B, and 11 - Type C) separated by device type.

We designed and fabricated three additional cantilever geometries (Fig. S4a): Type A the narrowest with the longest overhang; Type B a middle width with a slightly shorter overhang; and Type C the widest device with the shortest overhang (the device type reported in the main paper). Table S1 shows the overhangs and widths of the three types. We characterized a multitude of the 3 device types with DC and AC sweeps to get a sense of which geometry performs the best under piezo and electrostatic actuation.

**Table S1: Different cantilever geometries**

| Device | Overhang (μm) | Width (μm) | Waveguide loops |
|---|---|---|---|
| A | 221 | 342 | |
| B | 181 | 617 | |
| C (main text design) | 91 | 1232 | |

Fig. S4b and c show the results of the characterization. Generally, devices A and B performed worse than device C, i.e. had a higher $V_\pi$, both in AC and DC actuation strength. We hypothesize the longer overhangs of A and B caused the cantilever to begin in the contact (likely clamped) regime upon release, skipping the most effective operating regime of the pinned-to-clamped buckling transition. The contact starting point also increases mechanical losses for piezoelectric actuation, all of which show larger piezo $V_\pi$ across the board. However, the resonance shift as capacitive voltage increases is still seen in all three devices (likely due to length changes in the clamped regime). Fig. S4d also shows that generally device C is more consistent in performance than devices A and B. We attribute the differences to the undercutting process of fabricating these devices being more consistent for shorter overhangs as opposed to longer overhangs.